\documentclass[floatfix,nobalancelastpage,twocolumn,english]{revtex4} 
\usepackage{graphicx}
\usepackage{dcolumn}
\usepackage{bm}

\begin{document}

\preprint{APS/123-QED}

\title{Time-Resolved characterization of InAsP/InP quantum dots emitting in the C-band telecommunication window}

\author{R. Hostein}
\author{A. Michon}%
\author{G. Beaudoin}
\author{N. Gogneau}
\author{G. Patriache}
\author{J.-Y. Marzin}
\author{I. Robert-Philip}
\author{I. Sagnes}
\author{A. Beveratos}
\affiliation{%
Laboratoire de Photonique et Nanostructures LPN-CNRS UPR-20 \\
Route de Nozay. 91460 Marcoussis, France
}%

\date{\today}

\begin{abstract}
The dynamic response of InAsP quantum dots grown on InP(001)
substrates by low-pressure Metalorganic Vapor Phase Epitaxy emitting around 1.55 $\mu$m, is investigated by means of
time-resolved microphotoluminescence as a function of 
temperature. Exciton lifetime steadily increases from 1 ns at low temperature to reach 4 ns at 300K while the integrated photoluminescence intensity decreases only by a factor of 2/3. These characteristics give evidence that such
InAsP/InP quantum dots provide a strong carrier confinement even
at room temperature and that their dynamic response is not
affected by thermally activated non-radiative recombination up to
room temperature.
\end{abstract}

\maketitle

In recent years, self-assembled semiconductor quantum dots (QDs)
have been considered for utilization in quantum information
processing (such as quantum communication \cite{Michler2000,
Santori2002, Stevenson2006}), in addition to their
conventional applications in optoelectronics, for example.
However, optoelectronics and the implementation of practical
quantum computation networks require mostly the use of QDs
emitting in the telecommunications C-band (1.53 - 1.56 $\mu$m).
Emission at wavelengths around 1.55 $\mu$m has already been
achieved from QDs grown on a GaAs substrate by means of different
techniques, for instance by unusually low temperature growth
\cite{Maximov1999} or epitaxial growth on metamorphic buffer
layers  \cite{Ledentsov2003, Xin2003}. Yet, one of the most
attractive material combinations for fabricating QDs emitting in
the C-band is InAs on InP substrate. Most of the studies, involving
high-vacuum growth techniques such as molecular beam epitaxy,
\cite{Gonzalez2000, Gendry2004} 
of InAs QDs on (001)-oriented InP substrates,
demonstrated the formation of strongly elongated nanostructures
(namely, ``quantum dashes'' or ``quantum sticks'') rather than
three-dimensional QDs. To circumvent this difficulty, a few groups
have used (113)B-oriented InP substrates \cite{Frechengues1999},
which are however not compatible with the standard processes used
in the fabrication of optoelectronic devices. Metalorganic vapor
phase epitaxy (MOVPE) has also been used for the growth of
InAs/InP(001) QDs. Various studies have shown that MOVPE allows
one to obtain QDs rather than quantum dashes \cite{Marchand1997,
Michon2005}. Recently, microphotoluminescence signals around 1.5
$\mu$m evidenced sharp spectral features corresponding to
radiative transitions from trapped single electron-hole pairs in
such InAs/InP QDs grown by MOVPE \cite{Takemoto2004, Miyazawa2005,
Saint-Girons2006}. Conversely, not much work has been done on the
dynamic response of such MOVPE-grown dots and its dependence as a
function of the temperature \cite{Haiping1995}. Yet, measurements of this dynamic
response and the impact of thermally activated non-radiative
mechanisms are quite important to assess the potential of such QDs
devices for example for high-speed direct modulation or efficient
generation of quantum states of light at high temperature (more
than 77 K). This study forms the basis of this letter. The
dependence of the large measured radiative lifetime and the small
decrease of the integrated intensity as a function of temperature
demonstrate the high structural quality of these QDs,
which offer strong carrier confinement: non radiative
recombination inside the QDs and carrier evaporation to the
wetting layer are not critical to the operation of such
long-wavelength QD-based devices up to 300 K.

Samples were grown in a vertical-reactor low-pressure MOVPE system
using hydrogen as the carrier gas and standard precursors [arsine, phosphine, and trimethylindium]
\cite{Michon2005}. The quantum dots are grown on a thick InP
buffer layer deposited on an exactly (001)-oriented InP substrate. The QDs
are formed at a lowered temperature of 510 $^\circ$C by depositing 6.3 monolayers (ML) of InAsP at growth rate of 0.36 ML.s$^{-1}$ and under a phosphine/arsine flow ratio of 30.
Finally, a 63 nm thick InP capping layer is grown over the QDs at a rate of 0.2 ML.s$^{-1}$. Such a growth sequence
leads to the formation of InAsP QDs with an average height of 3.8 nm
and a density of 15x$10^9$ cm$^{-2}$, measured by \underline{T}ransmission \underline{E}lectron \underline{M}icroscopy (TEM) experiments \cite{Michon2008}. 
Time-resolved microphotoluminescence characterization of samples was obtained under pulsed excitation with a 5 ps pulse width
at 80 MHz delivered by a Ti:Sa laser emitting at 840 nm. The
excitation pulses were focused on the samples by a microscope objective (Numerical Aperture (NA) = 0.4). The excitation spot spreads
over roughly 5 $\mu$m on the sample, corresponding to the
excitation of approximatily 2.9x10$^3$ quantum dots. The QDs
luminescence is collected by the same microscope objective and
separated from the pumping laser by means of a dichroic mirror and
an antireflection coated silicon filter. The spontaneous
emission is spectrally dispersed by a 0.5 m spectrometer and
detected either by a cooled InGaAs photodiode array (Roper
Scientific) or a time-resolved single photon counter (SCONTEL) with a time
resolution of 50 ps, a measured quantum efficiency of 3\% at 1.55
$\mu$m and dark count rates lower than 30 counts per second. The histogram of the time interval 
between the photon detection and the subsequent laser pulse is recorded by a LeCroy 8620A oscilloscope.

\begin{figure}[!h]
\includegraphics[scale=0.7]{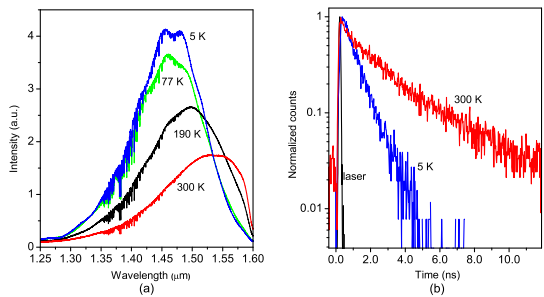}
\caption{\label{fig:Spectrum} (a) Emission spectra of the ground
states of InAsP/InP QDs under pulsed excitation as a function of
temperature. (b) Typical lifetime decay curves for
QDs at 5 K (blue) and room temperature (red). }
\end{figure}

All the following experiments are performed under low excitation power (of the order of 200 W/cm$^2$), so that only the fundamental transition of the QDs (one trapped electron-hole pair per dot at most) contributes to the optical emitted spectrum at low temperature (5 K). For each temperature a photoluminescence spectrum is recorded (fig. \ref{fig:Spectrum}(a)) and emission lifetime is measured (fig. \ref{fig:Spectrum}(b)). Emission spectra show excess noise in the 1350-1410 nm region arising from water absorption in air. Due to the QD's composition (InAs$_{0.82}$P$_{0.18}$) \cite{Michon2008}, leading to a smaller energy difference between subsequent monolayers, it is difficult to distinguish in such spectra different QD families; yet, it is possible to fit the photoluminescence spectra with a good fidelity by considering an 8-modal distribution of dot height. The low-temperature large inhomogeneous linewidth of the spectra is mainly due to height dispersion of the dots \cite{Michon2006}. It broadens with temperature due to the thermally activated population of the first excited states in the QDs, while the photoluminescence peak intensity decreases. The abrupt decrease of the photoluminescence around 1600 nm (775 meV) arises from the cutoff of the InGaAs detector array. Due to the redshift of the fundamental transition with temperature, lifetime measurements were performed at the maximum of the distribution probing the same quantum dot family (same height in monolayers) \cite{Fiore2000}.

\begin{figure}[!h]
\includegraphics[scale=0.8]{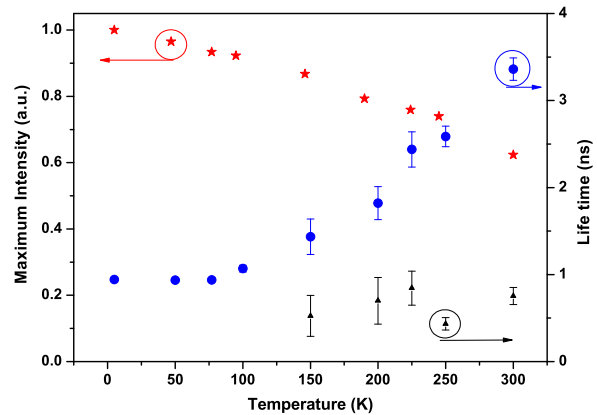}
\caption{\label{fig:LifeTimevsT} Measured lifetime (circles) and
photoluminescence peak intensity (stars) of the QDs as a function
of temperature. At temperatures above 200 K, the decay curves are
fitted by a biexponential decay with a fast (triangles) and a long
decay time (circles).}
\end{figure}

The
measured rise time of the order of 100 ps is limited by the timing
resolution of the set-up and does not correspond to the
carriers capture time in the QDs. At temperatures below 200 K, the experimental data can be fitted by a monoexponential decay, with a
constant decay time of the order of 1 ns for $T<$77 K (see Figure
\ref{fig:LifeTimevsT} ). Above 77 K and up to 200 K, the measured
lifetime increases from 1 ns to 1.8 ns. At higher temperatures
($\geq$ 200K), the data are well fitted by a
bi-exponential decay (see fig. \ref{fig:Spectrum}(b)) with a short
decay time of the order of 800 ps and a long decay time of the
order of few ns, plotted in fig.
\ref{fig:LifeTimevsT}. We only observe a small contribution of the
short decay photoluminescence, leading to large error bars, and it seems to display a flat
dependence as a function of temperature, while the long decay time
increases from 1.8 ns to 4.2 ns. We attribute the short decay to
the emission of thermally excited states (such as an electron-hole
pair on a p shell) of lower energy QDs and the long decay time to
the emission of the ground states of the QDs at the maximum of the
dots distribution. This hypothesis is corroborated by lifetime measurements under higher
excitation power (not shown here).

The constant lifetime value at low temperature
(here 1 ns up to 77 K) and the subsequent increase with
temperature (here from 1 ns to 4.2 ns for temperatures ranging
from 77 K to 300 K) has already been observed on InAs/GaAs quantum
dots \cite{Wang1994, Haiping1995, Fiore2000}. This increase is due
to the thermal dissociation of the exciton and the excitation of
the carriers to trapped excited states. Such states can be either
a hole or an electron in the p-shell, while the electron (or hole)
stays in the s-shell. These states are non-radiative and the
electron (or the hole) has to fall back to the s-shell in order to
emit a photon, which induces an increase of the measured
lifetime with no change on the total number of emitted photons.
These lifetime measurements suggest that the optical properties of
these dots are little affected by thermally activated
non-radiative mechanisms, which is corroborated by measurements of
the integrated photoluminescence intensity as a function of
temperature.  Although the current experimental setup does not allow
to deduce directly such value due to the cutoff of the InGaAs
array, an approximate value can be inferred by fitting the spectra
with a gaussian curve on the high energy part of the distribution.
These inferred values are plotted in fig \ref{fig:LifeTimevsT}. A
slight decrease of the integrated photoluminescence intensity can
be observed as the temperature rises, together with an increase of
the lifetime. Since the repetition rate of the laser is 12 ns, one
can not neglect the impact of the long QDs lifetime of the order
of few ns, which will induce a decrease in the integrated
intensity. Without correction, this decrease between 5 K and 300 K
is of only a factor of one third at most.

The observed weak impact of non-radiative processes in such QDs
are in good agreement with previous theoretical studies on
InAs/InP quantum dots \cite{Gong2008},
indicating that the InAs/InP
system  provides a stronger hole confinement compared to its
InAs/GaAs counterpart, with an hole energy spacing ranging from 20
to 40 meV compared to less than 20 meV for InAs/GaAs\cite{Gong2008}. 
Electron energy spacing is higher than 50 meV for both quantum dots.
This feature could explain the fact that no significant
change in the exciton lifetime is observed below 100 K
($kT\simeq$8 meV). Above 100K, one could expect a decrease in the
integrated signal, since the p shell of the hole, whose
wavefunction spreads more significantly in the barriers and thus
is more sensitive to surrounding defects, is thermally populated.
While such a behavior is significantly observed in InAs/GaAs
quantum dots, we only observe a small decrease. 
The experimental result is in good agreement with the theoretical estimations. While for InAs/GaAs less than 85\%
of the hole wavefunction lies inside the quantum dot, it is estimated to be more than 97\% for InAs/InP QDs for any trapped excited state\cite{Gong2008}. We expect that the small phosphorous incorporation 
during growth should not change drastically the hole confinement.
Consequently, such InAsP/InP QDs should be less sensitive to any
non-radiative defects located in the surroundings.
%
%

In conclusion, we have measured the decay times and the integrated
photoluminescence intensity  as a function of temperature from the
ground states of InAsP/InP quantum dots grown by MOVPE emitting at
1.55 $\mu$m.  The increase of the lifetime and
the slight decrease of the integrated intensity evidence the outstanding
structural quality of the QDs and that their
luminescence properties are not degraded by any non-radiative
thermally activated mechanisms up to room temperature. Such high
quality quantum dots could be used as gain medium for the
implementation of semiconductor laser diodes or optical
amplifier operating at room temperature as well as for the
implementation of single photon source for quantum communications
operating at 77 K.

The authors acknowledge financial support from
Conseil R\'egional d'Ile-de-France under CRYPHO project and
fruitful discussions with P. Senellart and
J. Bloch.

\appendix

\pagebreak

\end{document}